\documentclass{elsarticle}
\usepackage{graphicx,bm}
\usepackage{amssymb,amsmath}

\begin{document}

\begin{frontmatter}

\title{Critical chemotactic collapse}

\author{Pavel M. Lushnikov}

\address{
Department of Mathematics and Statistics, University of New Mexico, Albuquerque, NM 87131, USA
}



\begin{abstract}
A Keller-Segel model describes macroscopic dynamics of bacterial colonies and biological cells as well as dynamics of a gas of self-gravitating Brownian particles. Bacteria secret chemical which attracts
other bacteria so that they move towards chemical gradient creating nonlocal attraction between bacteria.  If
bacterial (or Brownian particle) density exceeds a critical value then the density collapses (blows up) in a finite time which corresponds to bacterial aggregation or gravitational collapse.
Collapse in the Keller-Segel model  has striking qualitative similarities with a nonlinear Schrodinger equation including critical collapse in two dimensions and supercritical collapse in three dimensions. A self-similar solution near blow up point is studied in the critical two-dimensional case and it has a form of a rescaled steady state solution which contains a critical number of bacteria. Time dependence of scaling of that solution has square root scaling law with logarithmic modification.
\end{abstract}

\end{frontmatter}

\maketitle

\section{Introduction}

Formation of singularity in a finite time (blow up) is a quite general phenomenon in many nonlinear systems including self-focusing in nonlinear optics, plasmas, hydrodynamics, and collapse of Bose-Einstein condensate \cite{CaflischPapanicolaou1993,KuznetsovZakharov2007}.   Blow up is often accompanied by a dramatic contraction of the spatial extent of solution, which is called collapse \cite{KuznetsovZakharov2007}.
Collapse typically occurs when there is self-attraction in a nonlinear system. Here a collapse in a system of bacteria or biological cells is considered. Below we refer to bacteria and cell as synonyms.
Bacteria often communicate through  chemotaxis, when bacteria move towards chemical gradient of substance called chemoattractant.  Bacteria can also secrete the same chemoattractant which creates nonlocal attraction between them through secretion, diffusion and detection of chemoattractant by other bacteria of the same type. Bacteria are self-propelled and without chemotactic clue center of mass of each bacteria  typically experiences random walk. Motion of bacterial colonies is thus affected by permanent competition of random-walk-based diffusion and chemotaxis-based attraction. Macroscopically averaged motion of bacteria can be described by a Keller-Segel model (see e.g. [3-17]
and references therein):
\begin{eqnarray}\label{pottscontinuousKellerSegel}
\partial _t \rho &=&D\nabla^2 \rho-\nabla\big [k \rho\, \nabla c\big ], \\
\partial_tc&=&D_c\nabla^2 c +\alpha\, \rho,  \label{pottscontinuousKellerSegelc}
\end{eqnarray}
 where $\rho({\bf r},t)$ is the density of at spatial point ${\bf r}$ and time $t$, $c$ is the concentration of chemoattractant, $D$ is the diffusion coefficient of bacteria due to random walk, the coefficient  $k>0$ characterize strength of chemotaxis,  $D_c$ is the diffusion coefficient of chemoattractant and $\alpha$  is the production rate of chemoattractant by bacteria.

  If density of bacteria is initially low then diffusion typically exceeds attraction and density remains low. E.g., a typical time scale for evolution of  the a low density bacteria {\it Escherichia coli} distribution in a petri dish  is about a day \cite{BrennerLevitovBudrene1998} (see Figure 3A in Ref. \cite{BrennerLevitovBudrene1998}).
  If initial density is not very small so that attraction wins over diffusion then bacteria experience aggregation (see Figure 3B in Ref. \cite{BrennerLevitovBudrene1998}). A typical time scale of such aggregation in experiment on {\it Escherichia coli} \cite{BrennerLevitovBudrene1998} is several minutes so it has an explosive character compare to evolution of bacteria outside of aggregation area. A goal of this Letter is to study how aggregation occurs from moderate initial bacterial density (below we give exact criterion for formation of aggregation).  Aggregation corresponds to the collapse of bacterial density in
  the Keller-Segel model  (\ref{pottscontinuousKellerSegel}), (\ref{pottscontinuousKellerSegelc}). Aggregation is a first step to a formation of multicellular organisms and quite important in biological applications \cite{Ben-Jacob}.  Near singularity the Keller-Segel model is not applicable when typical distance between bacteria is about size of bacteria.  In that regime modification of the Keller-Segel equation was derived from microscopic stochastic dynamics of bacteria which prevents collapse due to excluded volume constraint (different bacteria cannot occupy the same volume) \cite{AlberChenGlimmLushnikov2006,AlberChenLushnikovNewman2007,LushnikovChenalberPRE2008}. Here  however the original Keller-Segel model without regularization is considered.

Typically diffusion of a
chemical is much faster than cell diffusion, i.e.  $D/D_c\ll 1.$ E.g.,
$D/D_c\sim 1/40-1/400$ for the cellular slime mold {\it
Dictyostelium} \cite{HoferSherrattMaini1995}, and  $D/D_c\sim 1/30$
for microglia cells and neutrophils \cite
{LucaChavez-RossEdelstein-KeshetMogilner2003,GrimaPRL2005}.  Thus   Eq. (\ref{pottscontinuousKellerSegelc}) evolves on a much smaller time scale than Eq. (\ref{pottscontinuousKellerSegel}) and one can neglect time derivative in  (\ref{pottscontinuousKellerSegelc})  which gives a reduced Keller-Segel equation (RKS)
\begin{equation}\label{pottscontinuousKellerSegelreduced}
\begin{split}
\partial _t \rho&=\nabla^2 \rho-\nabla\big [\rho\, \nabla c\big ], \\
\nabla^2 c& =- \rho,
\end{split}
\end{equation}
where we assumed that $D, \ D_c, \alpha$ and $k$ are constants and rewrote  all variables in dimensionless form as $t\to t_0 t,$ $r\to t_0^{1/2}D^{1/2}r$, $\rho\to (D_c/t_0\alpha k)\rho$, $c\to (D/k)c$ and $t_0$ is a typical timescale of the dynamics of $\rho$ in Eq. (\ref{pottscontinuousKellerSegel}).

Eq. (\ref{pottscontinuousKellerSegelreduced}) also describes a dynamics of a gas of self-gravitating Brownian particles which has applications in astrophysics \cite{SireChavanis2002,SireChavanis2008}. In that case the second Eq. in (\ref{pottscontinuousKellerSegelreduced}) is a Poisson equation for the gravitational potential $-c$ while $\rho$ is the gas density (all units are dimensionless). The first Eq.  in (\ref{pottscontinuousKellerSegelreduced}) is a Smoluchowski equation  for $\rho$ in the gravitational potential $-c$. Below we refer to $\rho$ and $c$ as the density of bacteria and the concentration of chemoattractant, respectively, but all results of this Letter are equally true for a gravitational collapse of the  gas of self-gravitating Brownian particles.

The main result of this work is that solution of RKS  (\ref{pottscontinuousKellerSegelreduced}) in a spatial dimension $D=2$ near collapse time $t_0$ in neighborhood  of spatial location of collapse   $r=0$  has the following cylindrically-symmetrical self-similar form
\begin{equation}\label{pselfsimilarfull}
\begin{split}
  \rho&=\frac{1}{L(t)^{2}}\frac{8}{(1+y^2)^2},   \\
  c&=-2 \ln (1+y^2),   \\
 y&=\frac{r}{L(t)},   \\
 L(t)&=2e^{-\frac{2+\gamma}{2}}\sqrt{t_0-t}e^{-\sqrt{-\frac{\ln(t_0-t)}{2}}}\left [1+O\left (\frac{\left (\ln{[-\ln{(t_0-t)}]}\right )^3}{\sqrt{-2\ln{(t_0-t)}}} \right ) \right ],
\end{split}
 \end{equation}
 where $\gamma=0.577216\ldots$ is the Euler's constant and $L(t)\equiv L$ is the time-dependent spatial width of solution (below we often omit argument of $L$ for brevity).
This result has striking similarities with critical collapse in a nonlinear Schr\"odinger equation  (NLS) [24-33].
 E.g., collapsing solution of NLS has a form of a rescaled ground state soliton solution of NLS while (\ref{pselfsimilarfull}) has a form of a rescaled steady state solution of (\ref{pottscontinuousKellerSegelreduced}). Total number of bacteria in a self-similar solution (\ref{pselfsimilarfull}) $N=\int \rho \, d{\bf r}=8\pi$ is conserved which is similar to conservation of the total number of particles in collapsing NLS solution. Scaling of collapsing NLS solution has the same leading order form  $(t_0-t)^{1/2}$ as in  (\ref{pselfsimilarfull}) but a correction for NLS has a well-known log-log form $L(t)=(2\pi)^{1/2}(t_0-t)^{1/2}/(\ln{[-\ln{(t_0-t)}]})^{1/2}$ which is different from an exponent-of-square-root-of-logarithm for (\ref{pselfsimilarfull}). The logarithmic corrections in both RKS and NLS result from nearly exact balance between linear and nonlinear terms (respectively balance between diffusion and attraction for RKS and balance between dispersion and nonlinear attraction in NLS).  The log-log scaling for NLS is quite challenging for numerical verification \cite{99FP}. E.g., it is difficult to distinguish numerically the log-log scaling from a power of log scaling. In contrast, the exponent-of-square-root-of-logarithm scaling  (\ref{pselfsimilarfull}) is much faster and gives sizable corrections even for a moderate increase of an amplitude of $\rho$. Similar scaling to (\ref{pselfsimilarfull}) was obtained in Refs.   \cite{HerreroVelazquez1996,VelazquezSIAMJApplMath2002} using a formal matched asymptotic expansion of RKS. The order of error in
(\ref{pselfsimilarfull}) is however quite different from both \cite{HerreroVelazquez1996} and \cite{VelazquezSIAMJApplMath2002} because of power 3 of the log-log factor. In Ref. \cite{VelazquezSIAMJApplMath2002} a power of the log-log term is 1 instead. Result of Ref. \cite{HerreroVelazquez1996} predicts $L(t)\propto \sqrt{t_0-t}e^{-\sqrt{-\frac{\ln(t_0-t)}{2}}} [-\ln{(t_0-t)}]^{(1/4)(-\ln{(t_0-t)})^{-1/2}}$ scaling which only partially agrees with  (\ref{pselfsimilarfull}). This difference is  not a big surprise because the formal matched asymptotic expansion of Refs. \cite{HerreroVelazquez1996} and \cite{VelazquezSIAMJApplMath2002} does not provide a good control of errors. Another competing scaling is $L(t)\propto \sqrt{t_0-t}e^{-\frac{1}{2}\sqrt{-\frac{\ln(t_0-t)\ln{[-\ln{(t_0-t)}]}}{2}}}$ which was obtained in Ref. \cite{SireChavanis2002} (see also Ref. \cite{SireChavanis2008} for more discussion on that scaling). It remains a challenge for a future numerical simulations to verify which of these predictions of $L(t)$ are correct. Previous simulations \cite{BettertonBrennerPRE2001,SireChavanis2002} showed existence of corrections to the leading order  scaling $L(t)\propto \sqrt{t_0-t}$ but were not able to resolve the logarithmic corrections.

Principle difference of the analysis of RKS  (\ref{pottscontinuousKellerSegelreduced}) and derivation of the scaling (\ref{pselfsimilarfull}) in this Letter  from all previous works is the use of ideas from the NLS collapse analysis. It allows to find of an appropriate gauge transformation which brings a linearization operator about solution (\ref{pselfsimilarfull}) to the self-adjoint form.

In Section \ref{section:collapse} we study general properties of collapse of RKS, prove collapse existence and explore multiple similarities with collapse in NLS.
Section \ref{section:collapse2D} focuses on a self-similar solution of 2D RKS. We  write the self-similar solution as a rescaled steady state solution in blow up variables. We choose  a gauge transformation to a new dependent variable for perturbations about the self-similar solution to bring a linearization operator about the self-similar solution to a self-adjoint form.
In section \ref{section:amplitudeexpansion} we expand  these perturbations into eigenfunctions of the self-adjoint linearization operator to derive a set of amplitude equations for coefficients of that expansion.
Compatibility  conditions to  have an adiabatic form of that expansion result in a closed differential equation (\ref{ataufull}).
In Section \ref{section:blowuprate} we solve the equation (\ref{ataufull}) to derive the scaling (\ref{pselfsimilarfull}).
In Appendix we analyze a spectrum of the linearization operator and approximate its eigenfunctions.

\section{Collapse in the Reduced Keller-Segel Equation}
\label{section:collapse}

RKS (\ref{pottscontinuousKellerSegelreduced}) can be written in a form of a conservation law
\begin{eqnarray}
\partial _t \rho&=&-\nabla\cdot \Gamma, \label{KellerSegelGamma}
\end{eqnarray}
where   $\Gamma$ is the flux of bacterial probability density given by
\begin{eqnarray}
 \Gamma=-\rho\nabla \Big [\ln \rho-c \Big ], \label{Gammadef1}
\end{eqnarray}
and $c({\bf r})$ is determined by the fundamental solution $E({\bf r},{\bf r}')$ of the Poisson equation. Here we consider a two-dimensional bacterial motion with
\begin{eqnarray}\label{cdef1poisson}
c({\bf r})= -\int E({\bf r},{\bf r}') \rho({\bf r}')d{\bf r}',  \quad
E({\bf r},{\bf r}')=\frac{1}{2\pi}\ln |{\bf r}-{\bf r}'|.
\end{eqnarray}
Eq. (\ref{cdef1poisson}) allows to rewrite RKS (\ref{pottscontinuousKellerSegelreduced}) as a closed integro-differential equation for $\rho$. Nonlocality of interaction is due to diffusion of chemoattractant.

Assuming decaying boundary conditions at infinity we obtain a conservation of a total number $N$ of bacteria:
\begin{eqnarray}\label{Ndef1section5p1}
N=\int \rho({\bf r}) d {\bf r}=const.
\end{eqnarray}
One can also define a Lyapunov functional
\begin{eqnarray}\label{Edef1}
{\mathcal E}=\int \Big [\rho({\bf r}) \ln \rho({\bf r})-\rho({\bf r}) -\frac{\rho({\bf r}) c({\bf r})}{2}\Big ]  d {\bf r},
\end{eqnarray}
and represent RKS (\ref{pottscontinuousKellerSegelreduced}) in a gradient form
\begin{eqnarray}\label{gradient1}
\partial_t \rho=\nabla \cdot \left (\rho\nabla \frac{\delta {\mathcal E}}{\delta \rho} \right ), \quad  \frac{\delta  {\mathcal E}}{\delta \rho}=\ln \rho-c,
\end{eqnarray}
where the Lyapunov functional ${\mathcal E}$ is a non-increasing function of time
\begin{eqnarray}\label{Et1}
 \frac{d {\mathcal E}}{d t}=-\int \frac{\Gamma^2}{\rho} d {\bf r}.
\end{eqnarray}
 ${\mathcal E}$ is conserved only for a steady state solutions with zero flux $\Gamma=0.$

Although RKS (\ref{pottscontinuousKellerSegelreduced}) is a gradient non-Hamiltonian system, it has many striking similarities with NLS \cite{Sulem1999}:
\begin{equation}
i\partial_t \psi+ \nabla^2 \psi +  |\psi|^{2} \psi =0.
\label{nls1kellersegel}
\end{equation}
NLS  can be written in a Hamiltonian form $i\partial_t \psi=\frac{\delta H}{\delta \psi^*}$ with the Hamiltonian $H=\int  \left[ |\nabla\psi|^2-\frac{|\psi|^4}{2}\right ]d{\bf r}$. NLS  has huge number of applications, e.g. it describes dynamics of Bose-Eistein condensate, self-focusing of light and propagation of nonlinear dispersive waves.
Solutions of both RKS and NLS exist globally in dimension one $D=1$. Dimension $D=2$ is critical for both RKS and NLS  with collapsing solutions existing for $D\ge 2$ \cite{BrennerConstantinKadanoff1999,ChiaoGarmireTownes1964,vlas1971,zakh1972,ZakharovKuznetsov1986,lush1995,Sulem1999}. Collapse in critical dimension $D=2$ is strong for both  RKS and NLS, i.e. a finite number of bacteria (particles) is trapped into collapsing spatial region. For supercritical case $D>2$, e.g. for $D=3$, collapse in both RKS and NLS is weak which means that collapse is so fast that particles (bacteria) cannot keep up with collapse rate and close to collapse time $t\to t_0$ a vanishing number of bacteria (particles) are trapped into collapsing region.

To prove existence of collapse in RKS one can use a positive-definite quantity $A= \int r^2 \rho d{\bf r}$,  which determines a mean square width of bacterial density distribution \cite{BilerWoyczinski1998,ChavanisSire2006}.
Vanishing of $A$ guarantees existence of collapse because of conservation of $N$.  Proof of  collapse existence in NLS (\ref{nls1kellersegel}) is based on a virial identity $\partial^2_t B=8H$  \cite{vlas1971,zakh1972} and vanishing of $B$, where
 $B=  \int r^2 |\psi|^2 d{\bf r}$ is the  positive-definite quantity.  $B$ is the analog of  $A$ in RKS because $|\psi|^2$ in quantum mechanical interpretation of NLS  is the probability density of particle number, i.e. the analog of $\rho$ in RKS. However, RKS is the non-Hamiltonian system and the direct analogy with a virial theorem for $B$  does not work. Instead one can calculate  time derivative of $A$ using Eqs. (\ref{pottscontinuousKellerSegelreduced}),(\ref{cdef1poisson}), integration by parts and vanishing boundary conditions at infinity which gives for  $D=2:$
\begin{equation}\label{Atkellersegel}
A_t =4N-\frac{1}{2\pi}\int 2{\bf r}\cdot ({\bf r}-{\bf r}')\frac{\rho({\bf r})\rho({\bf r}')}{ |{\bf r}-{\bf r}'|^2}d{\bf r}d{\bf r}'=4N-\frac{N^2}{2\pi},
\end{equation}
 where we also used symmetrization over ${\bf r}$ and ${\bf r}'$. One concludes from (\ref{Atkellersegel}) that $A_t<0$ for $N>N_c=8\pi$   and $A$ turns negative in a finite time which proves existence of collapse.
Existence of the critical number of bacteria $N_c=8\pi$ is another similarity with NLS, where the critical number of particles $N_{c,NLS}=\int  |\psi|^2 d{\bf r}\simeq 11.70\ldots$ Qualitative  difference between RKS and NLS (\ref{nls1kellersegel}) is that RKS is the integro-differential equation while NLS (\ref{nls1kellersegel}) is  a partial differential equation (PDE).  However, it was shown in Ref. \cite{ref.lushnikov2002} that the virial identity can be used for a proof of collapse existence in an integro-differential equation of NLS-type  which describes Bose-Einstein condensate with nonlocal dipole-dipole interaction.

\section{Self-similar collapsing solution of the 2D reduced Keller-Segel equation}
\label{section:collapse2D}

Below we consider critical case $D=2$ only. Both RKS and NLS are scale invariant in 2D which means that if $\rho({\bf r},t), \, c({\bf r},t)$ is a solution of RKS then $L^{-2}\rho({\bf r}/L ,t/L^2),$ $c({\bf r}/L ,t/L^2)$ is also a solution for any  $L(t)\equiv L=const>0.$ Similar property holds for NLS.  RKS has a static solution
\begin{equation}\label{pstaticsection5p1}
\begin{split}
& \rho_0=\frac{8}{(1+r^2)^2},  \\
 & c_0=-2 \ln (1+r^2),
 \end{split}
\end{equation}
with the critical number of bacteria $N(\rho_0)=8\pi,$ which is another striking similarity with the ground state soliton solution of NLS (\ref{nls1kellersegel})  also containing the critical number of particles   $N_{c,NLS}.$

 Near collapse time $t\to t_0$ and collapse location $r=0$ the RKS solution of $\rho({\bf r},t)$   approaches to a cylindrically-symmetric self-similar solution in a form of a rescaled stationary solution (\ref{nls1kellersegel}) with time-dependent scale $L:$
\begin{equation}\label{pselfsimilar}
\begin{split}
& \rho(r,t)=\frac{1}{L(t)^{2}}\rho_0\left ( \frac{r}{L(t)} \right),   \\
 & c(r,t)=c_0(  \frac{r}{L(t)}),
  \end{split}
 \end{equation}
 where $L(t)\to 0$ for $t \to t_0$.

 We introduce a new dependent variable $m$ for cylindrically-symmetric solutions:
   \begin{equation}\label{mdef}
   m(r,t)=\frac{1}{2\pi}\int_{|r'|\le r}\rho({\bf r}',t)\ d{\bf r}',
   \end{equation}
which allows \cite{HerreroVelazquez1996} to rewrite RKS as a closed equation for  $m:$
   \begin{equation}\label{meq1}
\partial_t m=r\partial_r r^{-1}\partial_r m+r^{-1} m\partial_r m.
 \end{equation}
 $m(r,t)$ in (\ref{meq1}) has a meaning of a number of bacteria inside a circle of radius $r$ (up to a factor $2\pi$). Boundary condition for $m$ at $r\to \infty$ is simply related to the total number of bacteria: $m|_{r=\infty}= N/(2\pi)$.
 In contrast to RKS, (\ref{meq1}) is PDE for $m$. This simplification is possible for cylindrically-symmetric solutions of RKS only.

The steady state solution (\ref{pstaticsection5p1}) of RKS takes the following form for $m:$
\begin{equation}\label{pstaticm0}
m_0=\frac{4r^2}{1+r^2},
\end{equation}
and the self-similar solution (\ref{pselfsimilar}) becomes
\begin{eqnarray}\label{mselfsimilar}
&&m_{selfsimilar}=\frac{4y^2}{1+y^2}, \nonumber \\
&&y=\frac{r}{L}.
\end{eqnarray}
Boundary condition at infinity gives the critical number of bacteria: \\
 $2\pi m_{selfsimilar}\big |_{y\to \infty}\to 8\pi=const$ which is also the indication that bacterial collapse is strong because the number of bacteria trapped into collapse is constant.

 Assuming a power law dependence $L(t)\sim (t_0-t)^\beta$ in the self-similar solution (\ref{mselfsimilar}) one concludes that all terms in Eq.  (\ref{meq1})  are of the same order provided  $\beta=1/2,$ which is similar to NLS. Like NLS, the self-similar solution   (\ref{mselfsimilar}) is not an exact solution of Eq. (\ref{meq1}) and it is necessary to consider the logarithmic correction to  $L(t)\sim (t_0-t)^{1/2}$: $L= (t_0-t)^{1/2}f(\ln{(t_0-t)})$, where $f(\ln{(t_0-t)})$ is the slow function compare with $(t_0-t)^{1/2}$. That slow function comes from nearly exact balance between linear and nonlinear terms of RKS (between diffusion and attraction) and allows to introduce a small parameter
\begin{equation}\label{adef1}
a = -L(t)\partial_t L(t)>0,
\end{equation}
which is a slow function of $(t_0-t)^{1/2}$ compare with  $L$.  A balance between linear and nonlinear terms of RKS becomes better with decrease of $a\to 0$ .

Based on the analogy with the critical NLS we introduce in Eq.  (\ref{meq1}) new independent "blowup" variables \cite{DLS2009}:
 \begin{equation}\label{taudef1}
 \begin{split}
& y=\frac{r}{L},  \\
& \tau=\int^t_0 \frac{dt'}{L(t')^2},
\end{split}
\end{equation}
which produces from Eq.  (\ref{meq1}) an  equation for a new unknown function  $\varphi(y,\tau)\equiv m(r,t):$
\begin{equation}
\partial_\tau
\varphi=y\partial_y y^{-1}\partial_y\varphi+y^{-1}\varphi\partial_y\varphi-a
y\partial_y\varphi, \label{eqn:NLHBlowup}
\end{equation}
where $a$ is given by (\ref{adef1}). Near collapse $L(t)\to 0$ which means that $\tau\to \infty,$ i.e. collapse occurs at infinity for the new time $\tau$.
The advantage of moving to blow up variables is that the function $\varphi$ has bounded derivatives and the collapse time $t_0$ is eliminated from consideration (it is mapped to $\tau\rightarrow\infty$).

Because we assume that $a$ is a slow function, it is natural to look at solutions of Eq. (\ref{eqn:NLHBlowup}) in an adiabatic approximation when one can neglect $\tau$-derivative in the left hand side (lhs)  of Eq. (\ref{eqn:NLHBlowup}). Then, assuming that $|a|\ll 1$, one can expand a solution of  (\ref{eqn:NLHBlowup}) in powers of $a$ starting from (\ref{mselfsimilar}) in the power zero. However, term $-a
y\partial_y\varphi$ grows with $y$ and violates that expansion for large $y$. So the adiabatic approximation can only work locally for not very large $y$ which is also familiar in analysis of  collapse in NLS.
This however does not create big problems because for $t\to t_0$ it is important to take care of what is happening in the neighborhood of collapse point $y=0$ where the term $-a
y\partial_y\varphi$ is indeed a small correction. Behavior of the solution at large $y$ does not affect the self-similar solution near a collapse time.

It is convenient to present a general solution of   Eq. (\ref{eqn:NLHBlowup}) in the following form \cite{DLS2009}:
 \begin{equation}\label{vdef1}
\varphi(y,\tau)=\frac{4y^2}{1+y^2}+e^{\frac{a}{4}y^2}\frac{y^2}{y^2+1}v(y,\tau),
\end{equation}
where $v(y,\tau)$ is responsible for all corrections with respect to the self-similar solution  (\ref{mselfsimilar}). Substitution of  (\ref{vdef1}) into (\ref{eqn:NLHBlowup}) gives the following Eq.:
\begin{equation} \label{veq1}
\partial_\tau v+\hat {\mathcal{L}}_av=F.
\end{equation}
Here
\begin{equation}\label{Ldef01}
\hat{\mathcal{L}}_a=-\frac{1}{y^3}\partial_y y^3\partial_y
-\frac{8}{(1+y^2)^2}+\big [\frac{a^2}{4}y^2-2a+\frac{2a}{1+y^2}
\big ].
\end{equation}
is a linear operator corresponding to a linearization of (\ref{eqn:NLHBlowup}) with respect to  (\ref{mselfsimilar}), and the term
\begin{equation} \label{Fdef1section5p1}
F=\frac{a_\tau}{4}y^2 v-\frac{8a}{y^2+1}e^{-a y^2/4}+\frac{ay^2
v^2 }{2(y^2+1)}e^{a y^2/4}+\frac{2 v^2 }{(y^2+1)^2}e^{a
y^2/4}+\frac{yv v' }{y^2+1}e^{a y^2/4}
\end{equation}
 is responsible for all other terms including nonlinear terms in $v$, an inhomogeneous and linear terms. Generally $F$ cannot be zero because     (\ref{mselfsimilar}) is not an exact solution of  (\ref{eqn:NLHBlowup}) for nonzero $a$. Eqs. (\ref{vdef1})-(\ref{Fdef1section5p1}) are equivalent to Eq.  (\ref{eqn:NLHBlowup})  because up to now we have not made any approximations.

 Advantage of the definition  (\ref{vdef1}) is that the operator ${\mathcal{L}}_a=-\frac{1}{y^3}\partial_y y^3\partial_y+V(y)$ has a form of a radially symmetric Schr\"odinger operator in spatial dimension four ($D=4$) with a potential
\begin{equation} \label{Vpotentialdef}
 V(y)=-\frac{8}{(1+y^2)^2}+\big [\frac{a^2}{4}y^2-2a+\frac{2a}{1+y^2}
\big ].
\end{equation}
It means that  ${\mathcal{L}}_a$  is the self-adjoint operator with a scalar product
\begin{equation} \label{scalarproduct1}
\langle \psi ,\phi \rangle =\int\limits^{\infty}_{0} \psi(y) \phi(y) \,  y^3 \, dy.
 \end{equation}
 The potential $V(y)\to \infty $ for $y\to \infty$ which ensures that   ${\mathcal{L}}_a$ has only discrete spectrum. This allows to expand arbitrary $v$ in a discrete set of eigenfunctions of  $\hat {\mathcal{L}}_a$:
\begin{equation} \label{vexp1}
v=c_1\psi_1+c_2\psi_2+c_3\psi_3+\ldots,
\end{equation}
where $\psi_j(y)$ are the eigenfunctions of $\hat {\mathcal{L}}_a$,
\begin{equation} \label{eigenvalfunc01}
\hat {\mathcal{L}}_a\psi_j=\lambda_j \psi_j
\end{equation}
and $\lambda_j$ are the respective eigenvalues.

Note that a use of the scalar product (\ref{scalarproduct1}) which corresponds to the radially symmetric Schr\"odinger operator in  $D=4$ is simply the auxiliary mathematical trick because the operator  $\hat {\mathcal{L}}_a$ is self-adjoint with that scalar product.  However, all solutions obtained below correspond to  RKS   (\ref{pottscontinuousKellerSegelreduced}) with $D=2.$

\section{Amplitude equations}
\label{section:amplitudeexpansion}

In this Section we
 derive a set of amplitude equations for $c_1, \ c_2, \ldots$ from (\ref{vexp1}) which solve Eq. (\ref{veq1}). We solve these amplitudes equations exploiting the fact that at the leading order in $a$ the  solution of   (\ref{eqn:NLHBlowup}) is given by (\ref{mselfsimilar})  which we used in definition of (\ref{vdef1}). We expand all expressions below in integer powers of small parameters  $a$ and $\frac{1}{\ln{\frac{1}{a}}}$ keeping a lowest nontrivial order of $a$.

We show in Appendix that one can approximate three lowest eigenfunctions $\psi_1, $ $\psi_2$ and $\psi_3$ of ${\mathcal{L}}_a$ as follows:
\begin{equation}\label{eigvector01tot}
\begin{split}
&\psi_1=\frac{8}{1+y^2}e^{-ay^2/4}, \\
&\psi_2=\frac{8}{1+y^2}\left(1+
\frac{a y^2}{2}- \frac{ ay^2}{2}\ln{(1+y^2)}\right )e^{-ay^2/4},  \\
&\psi_3=\frac{8}{1+y^2}\left(1+
a y^2\left [-\frac{\pi^2 \ln{\frac{1}{a}}}{\pi^2-12}+ \frac{-12+\pi^2(2+\gamma-\ln{2})}{\pi^2-12}\right ]  \right . \\
&    \left . \quad \quad +ay^2\ln{(1+y^2)}\frac{12}{\pi^2-12} +\frac{a^2y^4}{4} \left [\ln{\frac{1}{a}}-3-\gamma+\ln{2}-\frac{24}{\pi^2-12} \right ] \right ) e^{-ay^2/4},
\end{split}
\end{equation}
where $\gamma=0.577216\ldots$ is the Euler's constant.
The discrete spectrum of $\hat {\mathcal{L}}_a$ is given by (see Appendix and Ref. \cite{DeOvSi2008}):
\begin{equation}\label{lambtot}
\begin{split}
&\lambda_1=a\left (-2+\frac{2}{\ln{\frac{1}{a}}}+2(1+\gamma-\ln{
2})\frac{1}{(\ln{\frac{1}{a}})^2}\right )
+O\left (\frac{a}{(\ln{\frac{1}{a}})^3}\right )+O(a^2),
 \\
&\lambda_2=a\left (\frac{2}{\ln{\frac{1}{a}}}+2(2+\gamma-\ln{
2})\frac{1}{(\ln{\frac{1}{a}})^2}\right )
+O\left (\frac{a}{(\ln{\frac{1}{a}})^3}\right )+O(a^2),  \\
&\lambda_3=a\left (2+\frac{2}{\ln{\frac{1}{a}}}\right )
+O\left (\frac{a}{(\ln{\frac{1}{a}})^2}\right )+O(a^2),
\end{split}
\end{equation}
and for $j>3$ the eigenvalues are given by Eq.  (\ref{lambdana}).

We assume (based e.g. on numerical simulations in \cite{BrennerConstantinKadanoff1999,BettertonBrennerPRE2001}) that $a$ is the adiabatically slow function of $\tau$: $a_\tau\ll a^2$.
An expansion of all amplitude equations  in integer powers of $a$ and $\frac{1}{\ln{\frac{1}{a}}}$ leads to a conclusion that $\partial_\tau a \sim a^2 O(\frac{1}{\ln{\frac{1}{a}}})$.
We introduce a normalized function  $\tilde a_\tau=\partial_\tau a/a^2=O(1)$ which is $\propto \partial_\tau a$ and depends on  $\ln{\frac{1}{a}}$ only. A subscript  $\tau$  indicates that $\tilde a_\tau$ comes from  $\partial_\tau a$. This allows to write $\partial_\tau a$  as an expansion in inverse powers of $\ln{\frac{1}{a}}$:
\begin{equation} \label{atau}
\partial_\tau a=a^2\, \frac{1}{\ln{\frac{1}{a}}} \ \tilde a_\tau, \quad \tilde a_\tau=\tilde
a_\tau^{(0)}+\tilde a_\tau^{(1)}\frac{1}{\ln{\frac{1}{a}}}+O\left (\frac{1}{(\ln{\frac{1}{a}})^2}\right ),
\end{equation}
where the coefficients $a_\tau^{(0)}$ and $a_\tau^{(1)}$ are $O(1)$ and do not depend on $\tau$ in the adiabatic approximation.

Assume that initially the expansion coefficients $c_1, \, c_2, \ c_3, \ldots $ in  (\ref{vexp1})  are $O(1)$. We  substitute the expansion (\ref{vexp1}) of $v$ from (\ref{vdef1}) into (\ref{veq1}) and scalar
multiply (\ref{veq1}) onto $\psi_1$, $\psi_2$ e.t.c. (with the scalar product
(\ref{scalarproduct1})) to obtain a set of amplitude equations for $c_1, c_2, \ldots$.
 A series expansion of these amplitude equations into small $a$ using Eq. (\ref{atau}) results in
\begin{equation}\label{psi1scalarall}
\begin{split}
&\partial_\tau c_1+a-2ac_1+O\left (\frac{a}{\ln{\frac{1}{a}}}\right )=0,  \\
&\partial_\tau c_2+O\left (\frac{a}{\ln{\frac{1}{a}}}\right )=0,  \\
&\partial_\tau c_3+2ac_3+O\left (\frac{a}{\ln{\frac{1}{a}}}\right )=0,  \\
&\partial_\tau c_4+4ac_4+O\left (\frac{a}{\ln{\frac{1}{a}}}\right )=0,  \\
&\ldots
\end{split}
\end{equation}
 where the terms $2a(j-2),$ $j=1,2,3,4,\ldots$ originates from eigenvalues for $\psi_j$, respectively (see Eq.  (\ref{lambdana})), the term $a$ in the first equation comes from the scalar product of $\psi_1$ with the second term in  the right hand side (rhs) of (\ref{Fdef1section5p1}) and we used the orthogonality of the vectors $\psi_1, \, \psi_2, \, \psi_3, \ldots.$  One concludes from Eqs. (\ref{psi1scalarall}) that the coefficients $ c_3, \, c_4, \ldots $ initially decay exponentially until they reach the quasi-steady (adiabatic) state with
\begin{equation} \label{cqeqall}
  c_3, \, c_4, \ldots =O\left (\frac{a}{\ln{\frac{1}{a}}}\right ).
  \end{equation}

In the first equation of (\ref{psi1scalarall}) we assume that
\begin{equation} \label{cqeq1}
c_1=\frac{1}{2}+O\left (\frac{a}{\ln{\frac{1}{a}}}\right )
\end{equation}
to avoid exponential growth of $c_1$ in $\tau.$

We have now a freedom in  $c_2$ and we choose it to have $v\to 0$ for any $y$ as $a\to 0.$
According to (\ref{eigvector01tot}), $\psi_1(y)|_{y=0}=\psi_2(y)|_{y=0}=8$ so we have to set
\begin{equation} \label{cqeq2}
c_2=-\frac{1}{2}+O\left (\frac{a}{\ln{\frac{1}{a}}}\right ).
\end{equation}
In that case $c_1\psi_1+c_2\psi=O(a)$ for $y=O(1)$, i.e.  $v$ from (\ref{mselfsimilar}) vanishes with vanishing of $a$.

Eqs.(\ref{cqeqall}),(\ref{cqeq1}) and (\ref{cqeq2}) justify the adiabatic approximation which means that the coefficients $c_1, \, c_2, \, c_3, \, c_4, \ldots$ depend on $\tau$ through $a$ only and one can series expand them in inverse powers  of $\ln{\frac{1}{a}}$:
\begin{equation} \label{cexp1}
\begin{split}
& c_1=\ \ \frac{1}{2}+d_1^{(1)} \frac{1}{\ln{\frac{1}{a}}}+d_1^{(2)} \frac{1}{(\ln{\frac{1}{a}})^2}+O\left (\frac{1}{(\ln{\frac{1}{a}})^3}\right )+O(a),
\\
& c_2=-\frac{1}{2}+d_2^{(1)} \frac{1}{\ln{\frac{1}{a}}}+d_2^{(2)}\frac{1}{(\ln{\frac{1}{a}})^2}+O\left (\frac{1}{(\ln{\frac{1}{a}})^3}\right )+O(a),
\\
& c_3=\quad \quad \quad d_3^{(1)}  \frac{1}{\ln{\frac{1}{a}}}+d_3^{(2)} \frac{1}{(\ln{\frac{1}{a}})^2}+O\left (\frac{1}{(\ln{\frac{1}{a}})^3}\right )+O(a),
\\
& \ldots
\end{split}
\end{equation}
where the expansion coefficients $d_i^{(j)} =O(1)$ for any $i, \,j$ and they do not depend on $\tau$ in the adiabatic approximation.

It follows from (\ref{cexp1}) and (\ref{atau})  that
\begin{eqnarray} \label{ctauall}
\partial_\tau c_j= O \left  (\partial_\tau \frac{1}{\ln{\frac{1}{a}}}\right )=O\left (\frac{a}{(\ln{\frac{1}{a}})^3}\right ), \quad j=1,2,3,\ldots.
\end{eqnarray}

Similar to Eqs. (\ref{psi1scalarall}) one can write amplitude Eqs. from the scalar multiplication of (\ref{veq1}) by $\psi_1, \, \psi_2$ and $\psi_3$, respectively.  Using (\ref{veq1})-(\ref{Fdef1section5p1}),(\ref{scalarproduct1})-(\ref{eigvector01tot}),(\ref{atau}),(\ref{cexp1}), and series expanding all expressions in   inverse powers  of $\ln{\frac{1}{a}}$  give the following equations:
\begin{eqnarray} \label{c1tot1}
&&\partial_\tau c_1 + \frac{a}{\ln{\frac{1}{a}}} \left
[\frac{\tilde a_\tau^{(0)}}{2}-2d_1^{(1)}\right
]
+\frac{a}{(\ln{\frac{1}{a}})^2} \left
[\frac{\tilde a_\tau^{(1)}}{2}+2d_1^{(1)}-2d_1^{(2)}+2d_2^{(1)} \right  . \nonumber  \\
&&\qquad \qquad \qquad \qquad \qquad \qquad\left .
 -\tilde a_\tau^{(0)}d_2^{(1)}+2d_3^{(1)}\right
]
+O\left (\frac{a}{(\ln{\frac{1}{a}})^3}\right )=0,
\\
&& \partial_\tau c_2 + \frac{a}{\ln{\frac{1}{a}}} \left
[-1-\frac{\tilde a_\tau^{(0)}}{2}\right
]
 +\frac{a}{(\ln{\frac{1}{a}})^2} \left
[-1-\frac{\tilde a_\tau^{(1)}}{2}+2d_2^{(1)}
\right  .  \nonumber  \\
&&\qquad \qquad \qquad\left .
+\tilde a_\tau^{(0)}(d_2^{(1)}-2d_3^{(1)})-\gamma+\ln{2})\right
]+O\left (\frac{a}{(\ln{\frac{1}{a}})^3}\right )=0,
\label{c2tot1}\\
&& \partial_\tau c_3 +  \frac{a}{\ln{\frac{1}{a}}} \left [2d_3^{(1)}\right ]+O\left (\frac{a}{(\ln{\frac{1}{a}})^2}\right )=0. \label{c3tot1}
\end{eqnarray}
Here  terms $\partial_\tau c_1, \, \partial_\tau c_2$ and $\partial_\tau c_3$ can be
omitted because they are $O\left (\frac{a}{(\ln{\frac{1}{a}})^3}\right )$  according to Eq. (\ref{ctauall}).
Eqs. (\ref{c1tot1})-(\ref{c3tot1}) are nothing more than compatibility conditions which ensure that expansions (\ref{cexp1}) and (\ref{atau}) are correct so that $a$ is indeed the adiabatically slow variable.

It follows immediately from Eq. $(\ref{c3tot1})$ in the order $\frac{a}{\ln{\frac{1}{a}}}$  that
\begin{eqnarray} \label{d3p1value}
d_3^{(1)}=0
\end{eqnarray}
and from Eq. $(\ref{c2tot1})$ in the order $\frac{a}{\ln{\frac{1}{a}}}$ that
\begin{eqnarray}\label{atau1}
\tilde a_\tau^{(0)}=-2
\end{eqnarray}
and then from Eq. $(\ref{c1tot1})$ in the order $\frac{a}{\ln{\frac{1}{a}}}$ we obtain
\begin{eqnarray} \label{d1p1value}
d_1^{(1)}=-\frac{1}{2}.
\end{eqnarray}
Using Eqs. (\ref{c2tot1}) and (\ref{d3p1value})-(\ref{d1p1value}) we obtain in the order $\frac{a}{(\ln{\frac{1}{a}})^2}$ that
\begin{eqnarray}\label{atau2}
\tilde a_\tau^{(1)}=-2-2\gamma+2\ln{2}.
\end{eqnarray}
Eqs. (\ref{atau}),(\ref{atau1}) and (\ref{atau2}) result in a closed differential equation for $a$:
\begin{equation} \label{ataufull}
\begin{split}
& \frac{a_\tau}{a^2}=-\frac{2}{\ln{\frac{1}{a}}}
+\frac{M}{(\ln{\frac{1}{a}})^2}+O\left (\frac{1}{(\ln{\frac{1}{a}})^3}\right ),  \\
& M=-2-2\gamma+2\ln{2}.
\end{split}
\end{equation}
It follows from Eqs. (\ref{cexp1}), (\ref{d3p1value}) and (\ref{d1p1value}) that the expansion coefficients (\ref{vexp1}) are given by
\begin{equation} \label{cexp2}
\begin{split}
& c_1=\frac{1}{2}-\frac{1}{2\ln{\frac{1}{a}}}+O\left (\frac{1}{(\ln{\frac{1}{a}})^2}\right ),
\\
& c_2=-\frac{1}{2}+O\left (\frac{1}{\ln{\frac{1}{a}}}\right ),
\\
& c_3=O\left (\frac{1}{(\ln{\frac{1}{a}})^2}\right ),
\end{split}
\end{equation}
so that the coefficients $c_3, \, c_4, \ldots$, which correspond to positive eigenvalues, are of a lower order compare with $c_1$ and $c_2.$  One concludes that the self-similar solution (\ref{mselfsimilar}) is stable with respect  to radially-symmetric perturbations and the leading order corrections to it are determined by a linear combination $v\simeq c_1\psi_1+c_2\psi_2,$ where $c_1$ and $c_2$ are given by (\ref{cexp2}).  Typical dynamics of perturbations around the self-similar solution is shown in Figure 4 of Ref. \cite{BettertonBrennerPRE2001}.

\section{Blow-up rate of self-similar solution }
\label{section:blowuprate}

In this section we solve Eq. (\ref{ataufull}) to derive a blow-up rate (\ref{pselfsimilarfull}).
Integration of Eq. (\ref{ataufull}) gives
\begin{eqnarray} \label{phieq}
\frac{1}{a}\left (\ln{\frac{1}{a}}+\left [\frac{M}{2}-1\right ]+O\left (\frac{1}{\ln{\frac{1}{a}}}\right ) \right )=2\tau,
\end{eqnarray}
which  is a version of the Lambert W-function and can be solved assuming $\tau\gg 1$ by iterations for $a$ as follows:
\begin{eqnarray} \label{phieqsol}
\ln{\frac{1}{a}}=\ln{2\tau}-\ln{\ln{2\tau}}+\frac{\ln{\ln{2\tau}}}{\ln{2\tau}}-\frac{\frac{M}{2}-1}{\ln{2\tau}}+O\left (\frac{(\ln{\ln{2\tau}})^2}{(\ln{2\tau})^2}\right ).
\end{eqnarray}

Now we use Eqs. (\ref{adef1}) and (\ref{taudef1}) to write $a=-L\partial_t L=-(\partial_\tau L)/L$ and express $L$ through $a$ as
\begin{eqnarray} \label{aint}
-\ln{L}= \int^\tau a(\tau')d\tau',
\end{eqnarray}
where the lower limit of integration is irrelevant for $\tau \gg 1.$
Using (\ref{phieqsol}) and (\ref{aint}) we obtain that
\begin{eqnarray} \label{Ltau}
-\ln{L}=\frac{(\ln{2\tau})^2}{4}-\frac{(\ln{2\tau})\ln{\ln{2\tau}}}{2}+\frac{M}{4}\ln{2\tau}+O\left( (\ln{\ln{2\tau}})^3\right ).
\end{eqnarray}

We solve Eq.  (\ref{taudef1}) for $t$ as $t_0-t=\int^\infty_\tau L(t)^2(\tau')d\tau'$, where $L(\tau)$ is given by  (\ref{Ltau}) which results for $\tau\gg 1$ in
\begin{equation} \label{tt0int}
\begin{split}
& t_0-t= (1/2)\left  [q^{-1}+O(q^{-2}\ln{q})\right ] \exp{\left[ -\frac{q^2}{2}+q\ln{q}-\frac{M}{2}q+q+O\Big ((\ln{q})^3\Big )\right ]},        \\
& q=\ln{2\tau},
\end{split}
\end{equation}
where $t_0$ is the collapse time.
Solving Eq. (\ref{tt0int}) by iterations for $\tau \gg 1$ we find an explicit expression for the rescaled time
\begin{equation} \label{tausol}
\tau=2^{-2}\sqrt{-2\ln{(t_0-t)}}e^{\sqrt{-2\ln{(t_0-t)}} +2+\gamma}
 \left [1+O\left (\frac{\left (\ln{[-\ln{(t_0-t)}]}\right )^3}{\sqrt{-2\ln{(t_0-t)}}} \right ) \right ]
\end{equation}
and using $1/L(t)^2=\partial_t\tau$  we obtain the scaling of the self-similar solution (\ref{mselfsimilar}):
\begin{eqnarray}\label{scaling2}
L(t)=2e^{-\frac{2+\gamma}{2}}\sqrt{t_0-t}e^{-\sqrt{-\frac{\ln(t_0-t)}{2}}}\left [1+O\left (\frac{\left (\ln{[-\ln{(t_0-t)}]}\right )^3}{\sqrt{-2\ln{(t_0-t)}}} \right ) \right ].
\end{eqnarray}
A similar scaling was obtained in Refs.   \cite{HerreroVelazquez1996,VelazquezSIAMJApplMath2002} using a formal matched asymptotic expansion of RKS.
Using Eq. (\ref{scaling2}) and returning from the variable $m$ of Eqs. (\ref{mdef}) and (\ref{meq1}) to the variables $\rho$ and $c$ of RKS   (\ref{pottscontinuousKellerSegelreduced}) we obtain Eqs. (\ref{pselfsimilarfull}) which is  the main result of this work.

\appendix
\section{Eigenvalues and eigenfunctions of the linearization operator  $\hat{\mathcal{L}}_a$}

The goal of this Appendix is to calculate the eigenvectors $\psi_j$ and the eigenvalues $\psi_j$ of the operator $\hat{\mathcal{L}}_a$  (\ref{Ldef01}).
It is impossible to write an explicit expressions for them  through standard mathematical functions so we have to use approximations.

The operator $\hat{\mathcal{L}}_a$  (\ref{Ldef01}) is reduced to
\begin{equation}\label{L0def}
\hat{\mathcal{L}}_0=-\frac{1}{y^3}\partial_y y^3\partial_y
-\frac{8}{(1+y^2)^2}-2a,
\end{equation}
for  $y \ll a^{-1/3} $ which is  the Schr\"odinger operator  for shallow potential well in $D=4$. $\hat{\mathcal{L}}_0$  corresponds to the boundary of disappearance of discrete spectrum (if we replace the potential in (\ref{L0def}) by a potential $-\frac{8-\delta}{(1+y^2)^2}, \ \delta>0$ then discrete spectrum is absent). Eq. $\hat{\mathcal{L}}_0\psi=\lambda \psi$ has  only one (up to an arbitrary constant factor) bounded  solution
 \begin{equation}\label{psizero}
\psi_{0}=\frac{8}{1+y^2},
\end{equation}
which is a soft mode  of that operator: $\hat{\mathcal{L}}_0\psi_0=2a\psi_0$ because $a\ll 1$. That soft mode corresponds to the breaking of scale invariance of the steady state solution  (\ref{pstaticm0}) in Eq.  (\ref{meq1}) because of a nontrivial time dependence of $L(t):$ $\psi_0(y)=-\frac{1+y^2}{y^2}y\partial_y m_{0}(y)$.

For the opposite limit $y \gg a^{-1/3} $ the   operator $\hat{\mathcal{L}}_a$ (\ref{Ldef01})  is reduced to the the Schr\"odinger operator for the spherically symmetric harmonic potential in $D=4:$
\begin{equation}\label{Lharmonic}
\hat{\mathcal{L}}_h=-\frac{1}{y^3}\partial_y y^3\partial_y
+\frac{a^2}{4}y^2-2a.
\end{equation}
Solution of $\hat{\mathcal{L}}_h\psi=\lambda \psi$ which decays for $y\to \infty$ is given by
\begin{equation}\label{psilambda}
\psi_h=a e^{-ay^2/4}U(-\frac{\lambda}{2a},2,\frac{ay^2}{2}),
\end{equation}
where $U(a,b,x)$ is the  confluent hypergeometric function of the second kind \cite{AbramowitzStegun}. Here we use the factor $a$ in rhs of (\ref{psilambda}) to match asymptotic of eigenfunctions for large $y$ as will be seen below.
An asymptotic of $\psi_h$ for $y \gg a^{-1/2}$ is given by \cite{AbramowitzStegun}
\begin{equation}\label{psilambdasymp}
\psi_h \propto e^{-ay^2/4}a^{\lambda/2a+1}y^{\lambda/a}\left [1+O\left ( \frac{\lambda^2}{a^3y^2} \right )   \right ].
\end{equation}

Matching of perturbations about the asymptotics (\ref{psizero}) and (\ref{psilambda})  allows to find the eigenvalues of  $\hat{\mathcal{L}}_a$ (\ref{Ldef01}) \cite{DeOvSi2008}.
In this work for calculation of the scalar products in Section \ref{section:amplitudeexpansion} we need not only eigenvalues but eigenfunctions also.  So  instead of matching the asymptotics we use a variational approximation for the eigenvalues and the eigenvectors of $\hat{\mathcal{L}}_a$. Idea is to satisfy simultaneously the asymptotics (\ref{psizero}) and (\ref{psilambdasymp}) for optimal choice of trial eigenfunctions.
For the first eigenvector $\psi_1$ of $\hat{\mathcal{L}}_a$ we choose
\begin{eqnarray}\label{eigvector01a}&&
\psi_1=\frac{8}{1+y^2}e^{-ay^2/4}
\end{eqnarray}
and approximate the eigenvalue $\lambda_1$ as
\begin{eqnarray}\label{eigenvalue01a}&&
\lambda_1=\frac{\langle
\psi_1,\hat{\mathcal{L}}_a\psi_1 \rangle}{\langle
\psi_1,\psi_1 \rangle}.
\end{eqnarray}
Calculation of (\ref{eigenvalue01a}) assuming $a\ll 1$ gives
\begin{equation}\label{lambda1a}
\lambda_1=a\left (-2+\frac{2}{\ln{\frac{1}{a}}}+2(1+\gamma-\ln{
2})\frac{1}{(\ln{\frac{1}{a}})^2}\right )
+O\left (\frac{a}{(\ln{\frac{1}{a}})^3}\right )+O(a^2).
\end{equation}

Here we have two small parameters $a$ and $\ln{\frac{1}{a}}$. We keep a minimal number of terms which are necessary for derivation of $L(t)$ dependence in Sections \ref{section:amplitudeexpansion} and \ref{section:blowuprate}.
One concludes from comparison of (\ref{psilambdasymp}) with (\ref{eigvector01a})    that they have the same asymptotic $\psi_1\propto y^{-2}e^{-ay^2/4}$  for $y\gg a^{-1/2}$.
One can now  see a motivation of our choice of the extra factor $a$ in  (\ref{psilambda}) because  $\psi_1$ in (\ref{eigvector01a}) with such choice matches the asymptotics between
$y \ll a^{-1/3} $ and $y \gg a^{-1/3} $ not only in power of $y$ but also in power of $a$.

From an analogy with the spectrum of the spherically symmetric harmonic oscillator like (\ref{Lharmonic}) we expect that the eigenvalues of  $\hat{\mathcal{L}}_a$ (\ref{Ldef01}) are separated at the leading order by $2a$ from each other. So that $\lambda_2\simeq O\left (\frac{a}{\ln{\frac{1}{a}}}\right )$ and from (\ref{psilambdasymp}) one concludes that $\psi_2\propto a e^{-ay^2/4}$  for $y\gg a^{-1/2}$. In addition, function $U(a,b,x)$ has terms $\propto 1/x$ and $\propto \ln{x}$ for small $x$ which all together motivates the following choice of a trial function for the second eigenfunction $\psi_2$:
\begin{eqnarray}\label{eigvector02a}
&&\psi_2=\frac{8}{1+y^2}\left(1+
c_2^{(2)}a y^2+c_l^{(2)}a y^2 \ln{(1+y^2)}\right )e^{-ay^2/4},
\end{eqnarray}
where the trial constants $c_2^{(2)}$ and $c_l^{(2)}$ should be chosen from an orthogonality condition
\begin{equation}\label{psi1psiorthogonality}
\langle \psi_1,\psi_2 \rangle=0
\end{equation}
and  from a minimization of the second
eigenvalue
\begin{equation}\label{eigenvalue02}
\lambda_2=\min\limits_{c^{(2)}_2, \,c_l^{(2)}} \frac{\langle
\psi_2,\hat{\mathcal{L}}_a\psi_2 \rangle}{\langle \psi_2,\psi_2 \rangle}.
\end{equation}
Our choice of the term $\ln{(1+y^2)}$ instead of $\ln{y^2}$ in (\ref{eigvector02a}) was motivated  by the necessity to avoid a logarithmic singularity for all real $y$. In addition, a similar term occurs if one approximates the eigenfunctions of $\hat{\mathcal{L}}_a$ (\ref{Ldef01}) by a perturbation expansion near the soft mode (\ref{psizero}) \cite{DeOvSi2008}.

Eqs. (\ref{eigvector02a}) and (\ref{psi1psiorthogonality}) result in the following expressions
\begin{equation}\label{cforeigvector02a}
\begin{array}{cccc}
& c^{(2)}_2&=&\frac{1}{2},  \\
& c^{(2)}_l&=&-\frac{1}{2}
\end{array}
\end{equation}
and
\begin{eqnarray}\label{lambda2a}
\lambda_2=a\left (\frac{2}{\ln{\frac{1}{a}}}+2(2+\gamma-\ln{
2})\frac{1}{(\ln{\frac{1}{a}})^2}\right )
+O\left (\frac{a}{(\ln{\frac{1}{a}})^3}\right )+O(a^2).
\end{eqnarray}
The coefficients $c^{(2)}_2$ and $c^{(2)}_l$ in (\ref{eigvector02a}) are chosen from a necessity to satisfy the orthogonality condition (\ref{psi1psiorthogonality}) up to the order $O\left (\frac{1}{\ln{\frac{1}{a}}}\right )$.
This choice is not unique because one can also use the minimization condition (\ref{eigenvalue02}). However, use of that condition would result in change of values of $\lambda_2$ in (\ref{lambda2a}) in terms of $O\left (\frac{a}{(\ln{\frac{1}{a}})^3}\right )$ which is beyond an order we need to calculate (\ref{pselfsimilarfull}). Thus our motivation for the specific choice  (\ref{eigvector02a}) is to have the most compact expressions for $c^{(2)}_2$ and $c^{(2)}_l$. It was checked that other possible choices of  $c^{(2)}_2$ and $c^{(2)}_l$ do not change both $\lambda_2$ in (\ref{lambda2a}) and final answer (\ref{pselfsimilarfull}) because they correspond to  higher order corrections $O\left (\frac{a}{(\ln{\frac{1}{a}})^3}\right )$.

From $\lambda_3\simeq 2a$ and  (\ref{psilambdasymp})  one concludes that $\psi_3\propto a^2 y^2 e^{-ay^2/4}$ for $y\gg a^{-1/2}$. Together with (\ref{eigvector02a}) that suggests the following choice for the third eigenfunction $\psi_3$:
\begin{equation}\label{eigvector03a}
\psi_3=\frac{8}{1+y^2}\left(1+
c_2^{(3)}a y^2+c_2^{(3)}a y^2 \ln{(1+y^2)}+c_4^{(3)}a^2y^4\right )e^{-ay^2/4},
\end{equation}
where the trial constants $c_2^{(3)}$, $c_l^{(3)}$ and $c_4^{(3)}$ should be chosen from two orthogonality conditions
\begin{equation}\label{psi3psiorthogonality}
\langle \psi_1,\psi_3 \rangle=0, \quad \langle \psi_2,\psi_3
\rangle=0,
\end{equation}
and from a minimization of the third
eigenvalue
\begin{equation}\label{eigenvalue03}
\lambda_3=\min\limits_{c^{(3)}_2, \,c^{(3)}_l, \, c^{(3)}_4}\frac{ \langle
\psi_3,{\mathcal{L}}_a\psi_3 \rangle}{\langle
\psi_3,\psi_3 \rangle}.
\end{equation}

Eqs. (\ref{eigvector03a}) and (\ref{psi3psiorthogonality}) result in the following expressions
\begin{eqnarray}\label{cforeigvector03a}
&& c^{(3)}_2=-\frac{\pi^2 \ln{\frac{1}{a}}}{\pi^2-12}+ \frac{-12+\pi^2(2+\gamma-\ln{2})}{\pi^2-12}, \nonumber \\
&& c^{(3)}_l=\frac{12}{\pi^2-12},  \\
&& c^{(3)}_4=\frac{1}{4} \left [\ln{\frac{1}{a}}-3-\gamma+\ln{2}-\frac{24}{\pi^2-12}\right ]\nonumber ,
\end{eqnarray}
and
\begin{eqnarray}\label{lambda3a}
\lambda_3=a\left (2+\frac{2}{\ln{\frac{1}{a}}}\right )
+O\left (\frac{a}{(\ln{\frac{1}{a}})^2}\right )+O(a^2).
\end{eqnarray}
Similar to $\psi_2$, the coefficients $c_2^{(3)}$, $c_l^{(3)}$ and $c_4^{(3)}$ in (\ref{eigvector03a}) are chosen from a necessity  to satisfy the orthogonality conditions (\ref{psi3psiorthogonality}) up to the order $O\left (\frac{1}{\ln{\frac{1}{a}}}\right )$.
This choice is not unique because one can also use the minimization condition (\ref{eigenvalue03}). However, use of that condition would result in change of values of $\lambda_3$ in (\ref{lambda2a}) in terms of
$O\left (\frac{a}{(\ln{\frac{1}{a}})^2}\right )$. But terms of that order  are not  needed to calculate (\ref{pselfsimilarfull}) in Sections \ref{section:amplitudeexpansion} and \ref{section:blowuprate}.

Eqs. (\ref{eigvector01a}),  (\ref{lambda1a}),  (\ref{eigvector02a}),  (\ref{cforeigvector02a}),  (\ref{lambda2a}),  (\ref{eigvector03a}),  (\ref{cforeigvector03a}),  (\ref{lambda3a}) result in Eqs. (\ref{eigvector01tot}) and (\ref{lambtot}). The eigenvalues (\ref{eigvector01tot}) are exactly the same as can be obtained from a matched asymptotic expansion   \cite{DeOvSi2008} which ensures accuracy of amplitude equations in Sections  \ref{section:amplitudeexpansion} and \ref{section:blowuprate}. As we mentioned above, choice of a trial functions is not unique. However we checked that different choices do not change results of Sections  \ref{section:amplitudeexpansion} and \ref{section:blowuprate} because they correspond to the higher order corrections. We also confirmed accuracy of the approximations of  the eigenfunctions  (\ref{eigvector01tot}) and the eigenvalues (\ref{lambtot}) by a direct numerical evaluation of the spectrum of $\hat{\mathcal{L}}_a$  (\ref{Ldef01}) for different values of $a$.

Other eigenvalues have the following  form \cite{DeOvSi2008}:
\begin{eqnarray}\label{lambdana}
\lambda_j=a\left (2(j-2)+\frac{2}{\ln{\frac{1}{a}}}\right )
+O\left (\frac{a}{(\ln{\frac{1}{a}})^2}\right )+O(a^2).
\end{eqnarray}
Corresponding eigenvectors can be obtained by an extension of the variational procedure described above through the inclusion of extra terms with the asymptotic $\psi_j\propto a^{j-1} y^{2(j-2)} e^{-ay^2/4}$ for $y\gg a^{-1/2}$.


The author thanks I.M. Sigal for attracting attention to this problem as well as for many  helpful discussions.


Work of P.L. was partially supported by NSF grants DMS 0719895 and
DMS 0807131.


\end{document}